\documentclass[10pt, conference, letterpaper]{IEEEtran}

\IEEEoverridecommandlockouts
\usepackage{cite}
\usepackage{amsmath,amssymb,amsfonts}
\usepackage{algpseudocode}
\usepackage{algorithm, algorithmicx, algpseudocode}
\usepackage{graphicx}
\usepackage{textcomp}
\usepackage{xcolor}
\usepackage{enumitem}

\begin{document}

\title{Dynamical ON-OFF Control with Trajectory Prediction for Multi-RIS Wireless Networks\\

}

\author{
\IEEEauthorblockN{Kaining Wang\IEEEauthorrefmark{1}, Bo Yang\IEEEauthorrefmark{1}, Yusheng Lei\IEEEauthorrefmark{1}, Zhiwen Yu\IEEEauthorrefmark{1}\IEEEauthorrefmark{2}, Xuelin Cao\IEEEauthorrefmark{3}, 
\\George C. Alexandropoulos\IEEEauthorrefmark{4},
Marco Di Renzo\IEEEauthorrefmark{5}\IEEEauthorrefmark{6}, and Chau Yuen\IEEEauthorrefmark{7} }

\IEEEauthorblockA{\IEEEauthorrefmark{1}School of Computer Science, Northwestern Polytechnical University, Xi'an, Shaanxi, 710129, China} 
\IEEEauthorblockA{\IEEEauthorrefmark{2}Harbin Engineering University, Harbin, Heilongjiang, 150001, China}
\IEEEauthorblockA{\IEEEauthorrefmark{3}School of Cyber Engineering, Xidian University, Xi'an, Shaanxi, 710071, China}
\IEEEauthorblockA{\IEEEauthorrefmark{4}Department of Informatics and Telecommunications, National and Kapodistrian University of Athens, 16122, Greece}
\IEEEauthorblockA{\IEEEauthorrefmark{5}Universit\'{e} Paris-Saclay, CNRS, CentraleSup\'{e}lec, Laboratoire des Signaux et Syst\`{e}mes, France}
\IEEEauthorblockA{\IEEEauthorrefmark{6}King's College
London, Centre for Telecommunications Research, Department of Engineering,  London, United Kingdom}
\IEEEauthorblockA{\IEEEauthorrefmark{7}School of Electrical and Electronics Engineering, Nanyang Technological University, Singapore}
}

\maketitle

\begin{abstract}
Reconfigurable intelligent surfaces (RISs) have demonstrated an unparalleled ability to reconfigure wireless environments by dynamically controlling the phase, amplitude, and polarization of impinging waves. However, as nearly passive reflective metasurfaces, RISs may not distinguish between desired and interference signals, which can lead to severe spectrum pollution and even affect performance negatively. In particular, in large-scale networks, the signal-to-interference-plus-noise ratio (SINR) at the receiving node can be degraded due to excessive interference reflected from the RIS. To overcome this fundamental limitation, we propose in this paper a trajectory prediction-based dynamical control algorithm (TPC) for anticipating RIS ON-OFF states sequence, integrating a long-short-term-memory (LSTM) scheme to predict user trajectories. In particular, through a codebook-based algorithm, the RIS controller adaptively coordinates the configuration of the RIS elements to maximize the received SINR. Our simulation results demonstrate the superiority of the proposed TPC method over various system settings.
\end{abstract}


\section{Introduction}
With the development of sixth-generation (6G) mobile communications, reconfigurable intelligent surfaces (RISs) have emerged as a groundbreaking technology to manipulate the wireless communication environment, significantly improving coverage and signal strength \cite{10596064,risRoadmap}, e.g., in aerial-terrestrial networks \cite{yangjsac}, vehicular networks \cite{TITS}, or underwater \cite{aRIS}. 
Although RISs can significantly enhance wireless communication systems, interference effects require detailed analysis. Recent standardization work\cite{noI2} and electromagnetic modeling studies\cite{noI1} have shown that, interference from the RIS direction can be negligible, especially in downlink system level scenarios with wide beam coverage and uniformly distributed users. In this case, narrow beamforming and spatial isolation can suppress the coupling effect of RISs.
However, very few existing studies have focused on the potential issue of spectrum pollution caused by `blind-reflection' of RISs. This is because RISs consist of nearly passive reflective elements incapable of differentiating between desired and interfering signals. 

\begin{figure*}
    \centering
    \includegraphics[width=1.7\columnwidth]{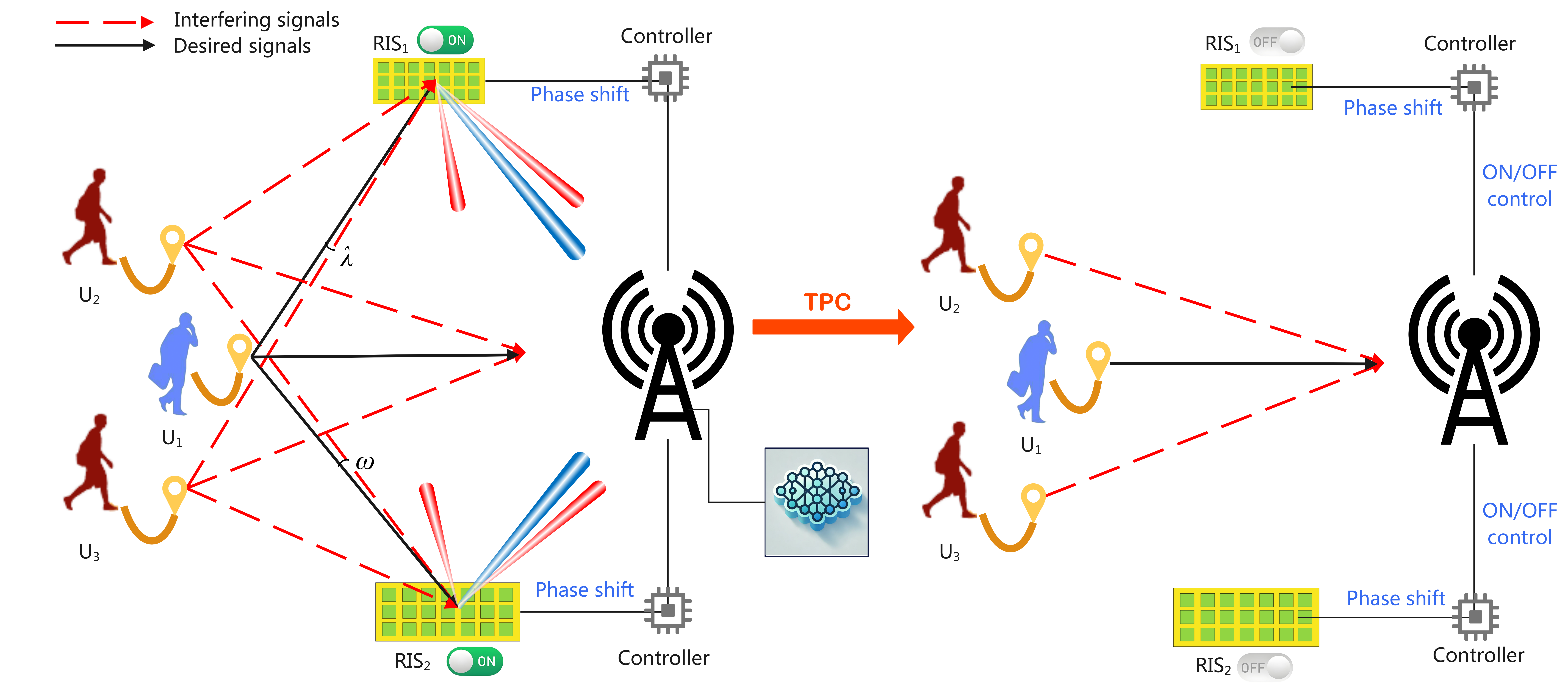}
    \caption{In the left figure, we show a conventional multi-user uplink RIS-assisted wireless communication system without TPC. Some interfering signals and desired signals are reflected to the BS. In the right figure, interfering signals approach the BS and are predicted by the BS. The BS sends the predicted control signal to the RIS based on the user trajectory. Each RIS controller can receive the control on/off signal and configure the RIS in real-time.}
    \label{fig:modelDesign}
\end{figure*}

A toy example is shown in Fig.~\ref{fig:modelDesign}, where the left side indicates the dense deployment scenario where the interfering users ($U_2$ and $U_3$) are very close to the desired user ($U_1$). In this context, the interfering users may locate on the straight line between the desired user and the RIS, e.g., $U_3$ is nearly located on the line between $U_1$ and RIS$_1$, the input angle meets $\lambda \approx  0^o$. Therefore, the overlapped radio frequency (RF) signals impinge on RIS$_1$, which cannot separate the signal but only reflect it, leading to unavoidable network interferences \cite{nointerference1,9864655,10729719}.

To address this problem, some research focuses on adapting to the complex changes in the channel environment and mitigating potential negative effects by adjusting the ON-OFF states of RIS elements. In \cite{tvt}, the authors dynamically configure these states to maximize the signal-to-interference-plus-noise ratio (SINR) by inferring the interfering signals from the desired signals through intelligent spectrum learning. In \cite{icc}, reinforcement learning is used to optimize the states of the RIS elements, aiming to reduce resource consumption. The authors of \cite{tcomm} achieve a balance between transmission rate and energy consumption by controlling the switches of the elements. Very recently, a neuro-evolution-based approach is presented for optimizing multiple-RIS-empowered wireless systems with discrete states. However, none of the above studies consider the potential impact caused by users' mobility, which is a typical feature of wireless networks.

In this paper we propose a trajectory prediction-based dynamical RIS ON-OFF control (TPC) scheme. Specifically, a long short-term memory (LSTM) model \cite{lstm} is deployed at the base station (BS) to predict the users' trajectory, impacting the distribution of the interfering signals. Based on this information, we further propose a distributed RIS binary control algorithm, enabling the BS to determine the ON-OFF states (active/inactive) of RISs in advance to maximize the SINR, as illustrated in the right side of Fig.~\ref{fig:modelDesign}. 

The rest of paper is organized as follows. Section \ref{s2} introduces the system model and problem formulation. Section \ref{s3} presents the proposed TPC scheme, followed by the simulation results in Section \ref{s4}.  Section \ref{s5} concludes the paper.



\section{System Model and Problem Formulation} \label{s2}
\subsection{RISs-aided Network Scenario}
We consider an RIS-assisted uplink communication system, which consists of one BS, $R$ RISs, and $U$ users, where $R \geq 1$ and $U \geq R$ usually hold. Let the set of RISs and users be $\mathcal{R}$ and $\mathcal{U}$, respectively. So, each RIS is denoted as RIS$_i, \forall i \in \mathcal{R}$, and each user is denoted as U$_u, \forall u \in \mathcal{U}$. As illustrated in Fig. \ref{fig:modelDesign}, we divide $U$ users into two subgroups: $U_L$ users who transmit the desired signals and the remaining $U_I$ users who are considered as interfering users, where $U_L+U_I=U$ holds. We assume that each RIS can serve only one user. 

Each RIS, denoted as RIS$_i$, has two states: ON and OFF. We introduce a binary variable, $v_i=\{0,1\}$, to indicate the state of RIS$_i$. Specifically, $v_i=1$ indicates that RIS$_i$ is ON so that the RIS$_i$ will reflect signals. Otherwise, $v_i=0$ indicates that RIS$_i$ is OFF, and in this case, no signal can be reflected. Moreover, we assume that each RIS has $N$ elements, and the phase can be continuously adjusted. The RIS controller can receive the signal from the BS to switch the ON-OFF state of elements.

\subsection{Channel Model}

We consider the product-distance PL model. Suppose the transmission power is $P$, and the small-scale fading is $g$. The direct link is assumed to be a Rayleigh fading channel, suitable for scattering-rich urban environments where no line-of-sight (LoS) path exists between the BS and the U$_u$. The path loss of the direct link can be expressed as
\begin{equation}
    \eta_u=Cd_{u}^{-\alpha },
\end{equation}
where $C=(\frac{c \sqrt{G_tG_r}}{4\pi f} )^2$ indicate the unit distance free space path loss that includes the c (speed of light), frequency ($f$), the transmit antenna gains ($G_t$), the receive antenna gains ($G_r$), and the free-space path index. $d_{u}$ is the distance between the U$_u$ and the BS. $\alpha$ is the path loss exponent. It is assumed that the horizontal distance of different channels is much larger than the vertical distance, and the heights of the BS, RIS, and user are ignored.

The path loss of the U$_u$-RIS$_i$-BS link can be expressed as
\begin{equation}
    \eta_{iu}=C(d_{i}d_{ui})^{-\alpha},
\end{equation}
where $d_{i}$ is the distance between the RIS$_i$ and BS, $\forall i\in \mathcal{R}$. $d_{ui}$ is the distance between the U$_u$ and RIS$_i$, $\forall u\in \mathcal{U}$. $C$ is the unit distance PL of the reflected link. The normalized small-scale fading of the RIS$_i$-BS and U$_u$-RIS$_i$ channels are denoted by $h_i$ and $h_{ui}$, respectively. 

The received signal from the desired user U$_l$ at the BS can be expressed as
\begin{equation}
    y_l=\left(\sqrt{\eta_l}g_l+\sum_{i=1}^{{R}}\sqrt{\eta_{li}}h_i\Phi_lh_{li}\right)\sqrt{P}s_l+I+\sigma,
\label{con:yk}
\end{equation}
where $\Phi_l$ denotes the diagonal reflection matrix of RIS-U$_l$, $s_l$ represents the signal sent by U$_l$, $\sigma$ is the additive white Gaussian noise (AWGN), $I$ represents the interference from all other RISs, i.e.,
\begin{equation}
    I=\sqrt{P} (\sum_{m=1}^{U_I}\sqrt{\eta_m}g_m+\sum_{i=1}^{{R}}\sqrt{\eta_{mi}} v_ih_i\Phi_mh_{mi}).
\end{equation}
  

\begin{figure*}
    \centering
    \includegraphics[width=1.24\columnwidth]{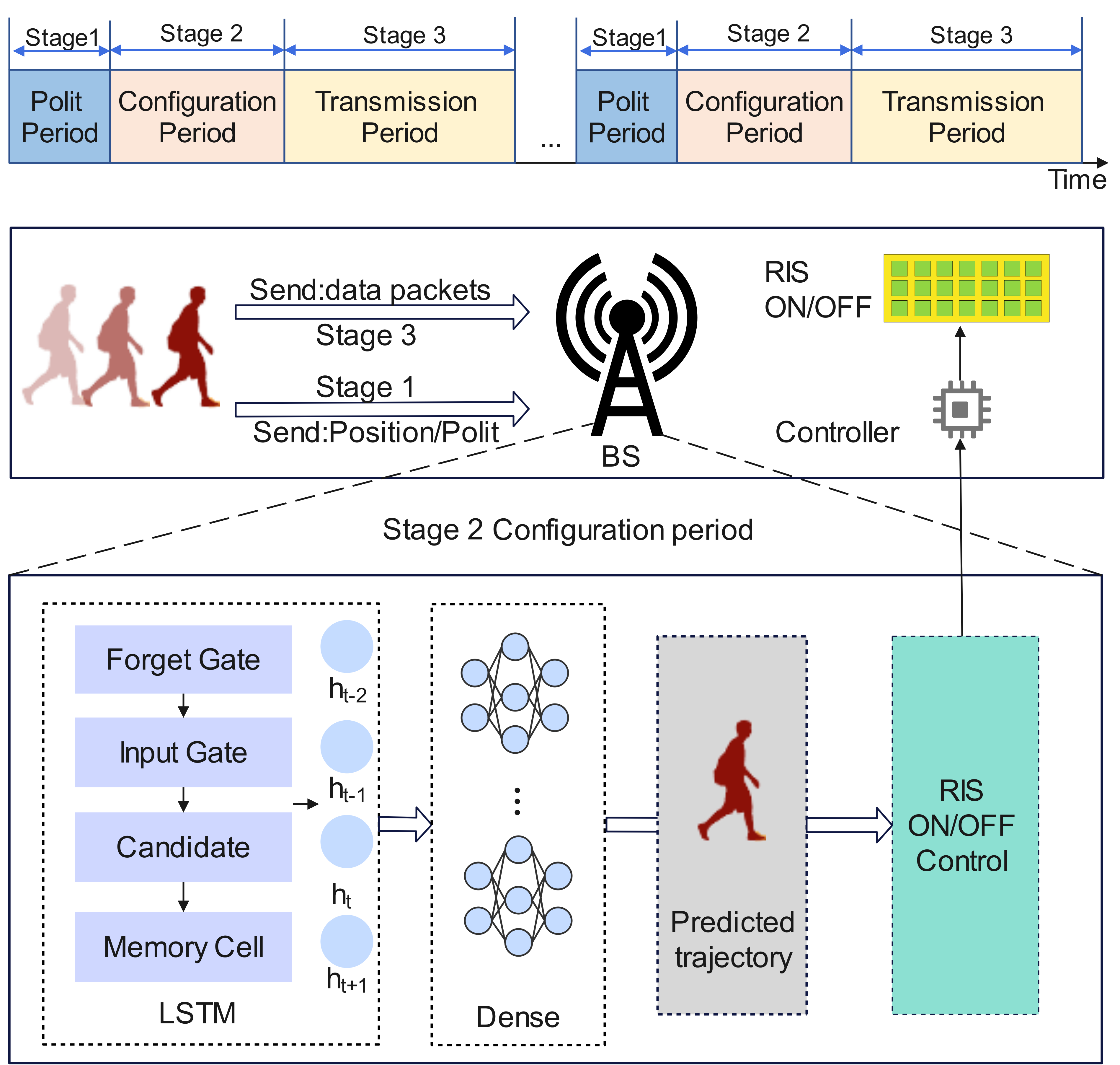}
    \caption{The proposed TPC algorithm structure. The input includes the history points of the pedestrian trajectories, and the output is the predicted coordinates.}
    \label{fig:algorithm}
\end{figure*}

\subsection{Problem Formulation and Analysis}
For the $l$th user (U$_l$), the SINR at the BS can be formulated as 
\begin{equation}\label{eq5}
	\gamma_l\!=\!\frac{P|\sqrt{\eta_l}g_l\!+\!\sum_{i=1}^{{R}}\sqrt{\eta_{li}}v_ih_i\Phi_ih_{li}|^2}{P|\sum_{m=1}^{U_I}\sqrt{\eta_m}g_m\!+\!\sum_{i=1}^{{R}}\sqrt{\eta_{mi}} v_ih_i\Phi_ih_{mi}|^2\!+\!\sigma^2}.
\end{equation}

In \eqref{eq5}, the numerator is the signal power of U$_l$, and the denominator includes the noise and interferences from all other RISs in the network.

We aim to optimize the phase shift matrix by optimizing the RIS binary activation vector (denoted as ${\mathbf V}=\{v_1,v_2,...,v_R\}$) and the RIS phase shift vector. Therefore, the optimization problem can be formulated as
\begin{equation}
\begin{aligned}
P_1:\quad & \max_{\mathbf V, \mathbf \Phi}\ \gamma_l \\
& \text{s.t. } v_i \in \{0, 1\}, && \forall i \in \mathcal{R},\\
& |e^{j\theta_n^i}| = 1, && \forall n \in [1, N],\ \forall i \in \mathcal{R}.
\end{aligned}
\end{equation}

We observe that the formulated optimization problem is a mixed-integer nonlinear program (MINLP), which is a class of mathematical optimization problems containing both discrete integer variables and continuous variables with at least one nonlinear component in the objective function or constraints. The optimization variables, reflection unit states, and continuous phase shifts form a hybrid discrete-continuous decision space.  To solve this problem, we propose the TPC algorithm, which is introduced in the following section.

\section{Proposed TPC Algorithm}\label{s3}
This section introduces a trajectory prediction method based on LSTM networks, which is designed to assist the BS in determining the user’s future location. We then present a trajectory prediction-assisted RIS control algorithm, illustrated in Fig.~\ref{fig:algorithm}. The proposed algorithm utilizes a well-trained trajectory prediction model at the BS to infer the user's position in near-real-time. Based on the inferred information, the optimal ON-OFF switching sequences and phase configurations can be obtained in advance to mitigate the interference problems.

\subsection{LSTM-based Trajectory Prediction}
Trajectory prediction involves forecasting a user's future coordinates based on past trajectories. Here, the LSTM model is utilized because it effectively captures long-term dependencies in sequential data, making it a popular choice for this task.

\subsubsection{Trajectory collection} 
In the context of wireless communication standards, such as 3GPP Release 18 \cite{R18} and Release 19 \cite{R19}, the BS is capable of obtaining and managing the users' location data effectively to comply with regulatory requirements (e.g., emergency calls), which necessitate a horizontal accuracy of less than or equal to $50$ meters, as well as for lawful interception. In addition, the Release 19, which is still in the pre-research phase and aims to support future 6G networks, suggests integrating an AI model in the physical layer. This model would directly predict the user's position using channel information, reducing the errors commonly associated with traditional algorithms.

Therefore, for our trajectory prediction task, the BS can effectively gather the users' historical locations. Based on this information, we can design and train the LSTM model to achieve trajectory prediction. Specifically, we use the trajectory dataset collected from a real environment in Beijing, which consists of a large number of users' latitude, longitude, and time information. We construct a compliant LSTM trajectory prediction model from these data, which can infer the switching sequences of the RIS in advance by predicting the positions of the users and the interferers.

\subsubsection{Offline training} 

The model is trained using collected users' trajectory information, and its structure is illustrated in Fig.~\ref{fig:algorithm}. Within this structure, the trajectory sequence is processed as a sliding window, and 64 independent trajectory segments are input at a time, each containing the latitude and longitude information of eight consecutive trajectory points. After being processed by a two-layer LSTM network, the model outputs the prediction results of the latitude and longitude coordinates of future trajectory points. The LSTM model can determine which historical information is still important and retain it, as well as introduce new content based on the relevance of the current input, effectively capturing long-term dependencies in the time series and reflecting the dynamic response to the input information. 

During the training process, the system employs the mean square error (MSE) as the loss function and uses the Adam optimizer for tuning. The LSTM trajectory prediction model is trained on TensorFlow using an AMD Ryzen 7 7700 8-Core Processor and an NVIDIA GeForce RTX 4060 Ti GPU.

\subsubsection{Online inference} 
The trained LSTM model begins by encoding the user's past trajectory. At the BS, feed-forward computations are performed to determine the user's position information for the upcoming frames. The feature data processed by the LSTM layer is then passed to a fully connected layer, which maps the hidden state vectors to the output space through a nonlinear transformation. This process predicts the pedestrian's coordinate position for the next moment. This layered processing mechanism preserves the correlation of features in the temporal dimension while ensuring accurate coordinate projection in the spatial dimension.

\subsection{RIS ON-OFF control}
After receiving the user's access request, the BS predicts the user's future trajectory. By implementing the RIS control algorithm at the BS, the binary ON-OFF state of the corresponding RIS can be determined, as shown in Fig.~\ref{fig:algorithm}.
\subsubsection{Codebook-based phase configuration}
To calculate the phase shifts at the RIS, we denote the channel state information (CSI) associated with the $i$th RIS as $\{ h_i, h_{il},h_{im},g_l, g_m\}$. The received SINR for the signal transmitted from user $U_l$ is determined based on the inferred coordinates of the users, which are obtained from the BS using the trained LSTM model, as expressed in (\ref{eq5}).

Based on the SINR, we first obtain the phase shifts of each RIS under the assumption $v_i = 1,\ \forall i \in \mathcal{R}$. Therefore, the original problem $P_1$ can be simplified as
\begin{equation}
\begin{aligned}
P_2:\quad & \max_{\mathbf \Phi} \ \gamma_l \\
\text{s.t.}\quad & v_i = 1,\ \forall i \in \mathcal{R}, \\
& |e^{j\theta_i^n}| = 1, \ \forall n \in [1, N],\ \forall i \in \mathcal{R}.
\end{aligned}
\end{equation}

The phase configuration adopts a predefined codebook approach \cite{codebook1,codebook}, where the codebook is defined as
\begin{equation}
\mathbf{\Phi} = \{\Phi_1,...,\Phi_i,\ldots,\Phi_R\},\end{equation}
where $\Phi_i = {\rm diag}(e^{j\theta_1^i},\ldots,e^{j\theta_N^i})$.

We suppose that the phase quantization bits are denoted as $b$ and let $\theta_n^i \in \{0,\frac{2\pi}{2^b},\ldots,\frac{2\pi(2^b-1)}{2^b}\}$. For the RIS$_i$, the optimal phase selection is performed as
\begin{equation}
    \Phi_i^* = {\rm argmax}_{\Phi \in \mathbf{\Phi}} \ \Gamma_i,
\end{equation}
where $
    \Gamma_i = \left| \sqrt{\eta_l}g_l + \sqrt{\eta_{li}}h_i\Phi_i h_{li} \right|^2 
$, $h_i$ and $h_{li}$ represent the channel gains of the links RIS$_i$-BS and U$_l$-RIS$_i$,  respectively.

Taking $b=2$ as an example, the switching states of the $N$ reflection units generate $2^N$ discrete configuration patterns, and each pattern needs to be optimized jointly with the $N^i$-dimensional continuous phase space, resulting in the dimensionality of the feasible solution space growing exponentially with the size of the RIS.

Based on the obtained optimal phase shifts $\mathbf{\Phi}^*$, we can further optimize the binary ON-OFF vector of RISs, i.e.,  $\mathbf{V}^*$. 

\subsubsection{RIS binary control algorithm}
If the RIS$_i$ is ON ($v_i = 1$), the signal sent by U$_u$ is reflected by RIS$_i$, so the received SINR at the BS via the RIS$_i$ is given by (\ref{eq5}).

When all RISs are OFF ($v_i = 0$), the received SINR at BS via the direct link is calculated as
\begin{equation}
	\gamma_l'=\frac{P|\sqrt{\eta_l}g_l|^2}{\sum_{m=1}^{U_I}P|\sqrt{\eta_{m}}g_m|^2+\sigma^2}.
\label{gammaD}
\end{equation}

To optimize system performance, the RIS$_i$ should be switched to `ON' when the condition $\gamma_l \geq \gamma_l'$ is met, as this helps improve signal quality. Conversely, in situations where the incident angles of the desired and interfering signals are relatively close, it is advisable to switch the RIS$_i$ to `OFF' to prevent significant interference, as illustrated in \textbf{Algorithm \ref{alg1}}.

\begin{algorithm}[thb]
\caption{RIS ON-OFF Control}\label{alg1}
\renewcommand{\algorithmicrequire}{\textbf{Input:}}
\renewcommand{\algorithmicensure}{\textbf{Output:}}
\begin{algorithmic}[1]
\Require Predicted coordinates of users, codebook $\mathcal{C}$
\Ensure Optimal ON-OFF  vector ${\mathbf V^*}\!=\! \{v_1^*,v_2^*,\ldots,v_R^*\}$
\State Initialize: $v_i = 1, \forall i \in \mathcal{R}$
\For{$i=1, i\leq R$}
    \State Calculate $\gamma_l$ and $\gamma_l'$ via (\ref{eq5}) and (\ref{gammaD}), respectively;
    \State Calculate $\Phi_i^*$ by solving problem $P_2$.
    \If{$\gamma_l$ \textless $\gamma_l'$} 
        \State $v_i^*=0$, i.e., turn the RIS$_i$ OFF.
    \Else
         \State $v_i^*=1$, i.e., keep the RIS$_i$ ON.        
    \EndIf
    \State  $ i\leftarrow i+1$;
    \State Update $v_i^*$ in $\mathbf{V^*}$.
\EndFor
\end{algorithmic}
\end{algorithm}

\subsubsection{Frame structure design}
Each time frame structure of the TPC algorithm involves three-stage periods:
\begin{itemize}
\item {Stage 1 - pilot period}: During the pilot phase, the user transmits the channel estimation pilot signal and the position reference signal at the same time. The channel state information reference signal is responsible for obtaining the instantaneous channel parameters. Meanwhile, the position reference signal extracts user position characteristics, such as time of arrival, longitude, and latitude, which are essential for reconstructing the user's trajectory.
\item {Stage 2 - configuration period}: Utilizing the historical trajectory data, the LSTM model can predict the user's position and assess the potential interference in advance. Following the steps outlined in \textbf{Algorithm \ref{alg1}}, the BS generates the RISs' ON-OFF switch sequence and selects the optimal phase configuration from a predefined codebook for the active RIS.
\item {Stage 3 - transmission period}: In the data transmission phase, the users transmit their data packets to the BS according to the configurations of RISs.

\end{itemize}


\section{Simulation Results} \label{s4}
To thoroughly evaluate the performance of the trajectory prediction model, this study uses the Geolife Trajectories dataset \cite{geolife} as the experimental foundation and selects several representative trajectory files as test samples. The proposed prediction framework is based on an LSTM model. The model inputs consist of continuous historical trajectory points over time steps, while the outputs are the future trajectory point coordinates corresponding to those inputs. To improve the generalization ability of the model, we selected $71$ trajectory files for training and testing. We used Mean Squared Error (MSE) as the primary loss function to optimize parameters in this spatial coordinate regression task. In the model evaluation phase, we used the Haversine formula to calculate the geographic distance error between the predicted trajectory points and the actual trajectory points. The mean error, measured in meters, served as the primary evaluation metric to assess the accuracy of the model's predictions. 

\subsection{Trajectory Prediction}
\begin{figure}
    \centering
    \includegraphics[width=1.02\columnwidth]{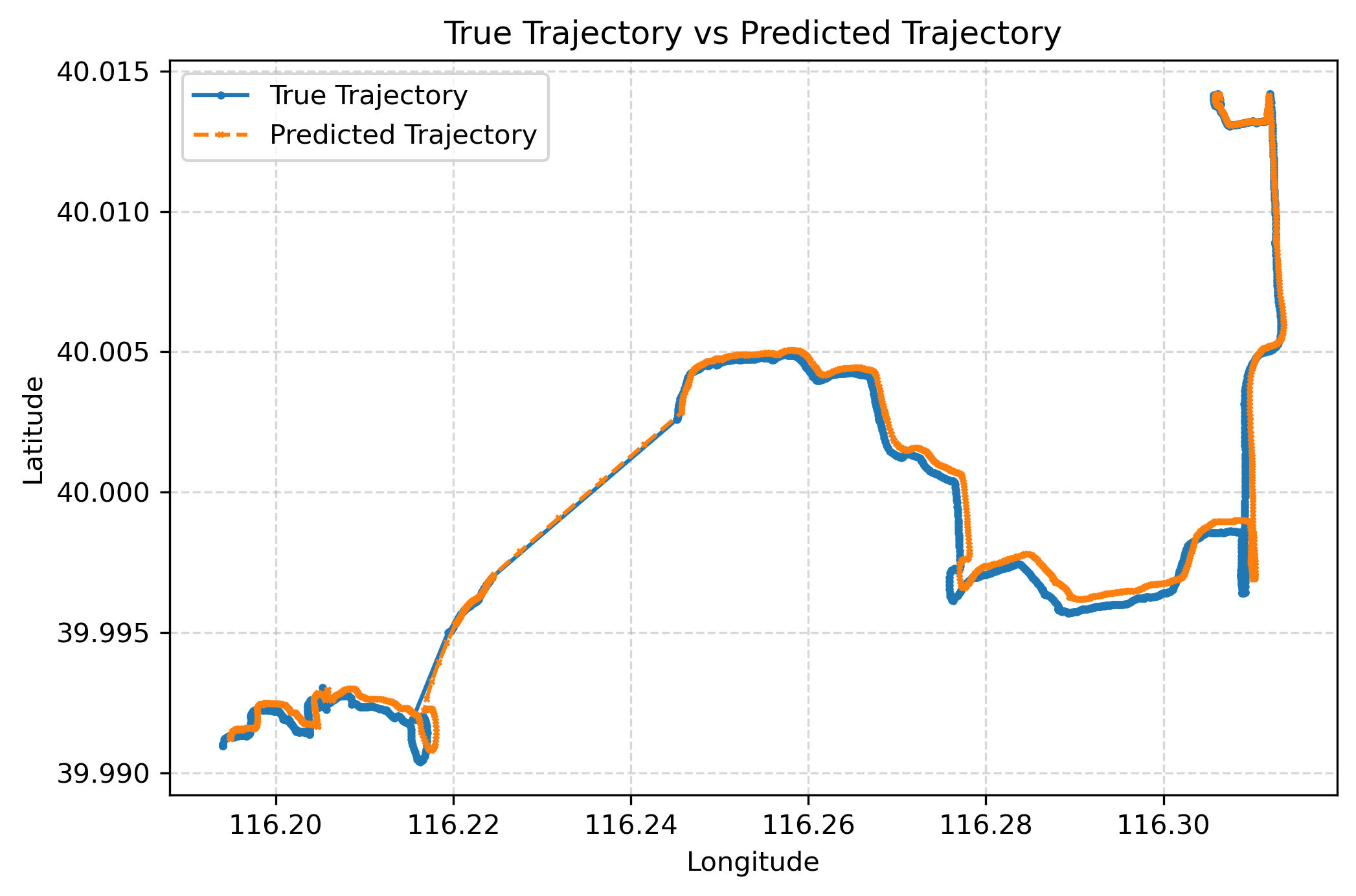}
    \caption{Comparison of real trajectory and predicted trajectory}
    \label{fig:result3}
\end{figure}

To verify the prediction performance, we specifically selected a representative test trajectory for visual analysis and used the trained LSTM model to perform trajectory prediction. As shown in Fig.~\ref{fig:result3}, we observe that the blue solid line represents the actual trajectory, while the orange dashed line depicts the predicted trajectory. From the visualization results, it is evident that the predicted trajectory closely aligns with the actual trajectory overall. Notably, at key turning points, along long straight segments, and in areas with rapid trajectory changes, the model maintains a stable tracking of the trajectory direction. 
The experimental results demonstrate that the prediction error of the model remains stable within ±$60$ meters for most test trajectories. This confirms the effectiveness of the LSTM model in capturing the spatiotemporal features of geographic trajectories. Additionally, these findings indicate that the designed LSTM model can effectively learn the overall evolutionary trends of the trajectories while also maintaining good performance in areas with frequent changes in trajectory.

Overall, the well-trained LSTM-based trajectory prediction model exhibits high prediction accuracy across various trajectory scenarios and thus can provide high-quality trajectory data for supporting subsequent wireless communication optimization tasks such as dynamic RISs ON-OFF switching control.


\subsection{Performance Evaluation}
\subsubsection{Simulation settings}
Unless otherwise specified, the simulation model includes one BS, $10$ RISs, one desired user, and $10$ interfering users. Each RIS consists of $600$ elements, and all RISs are evenly arranged in a circle around the BS, with a radius of $10$ meters. It is assumed that the reflected power from the RIS is inversely proportional to the incident angle of the signal. The target transmission power is $1$ watt, while the noise power is $1$ picowatt.

\begin{figure}
    \centering
    \includegraphics[width=1.1\columnwidth]{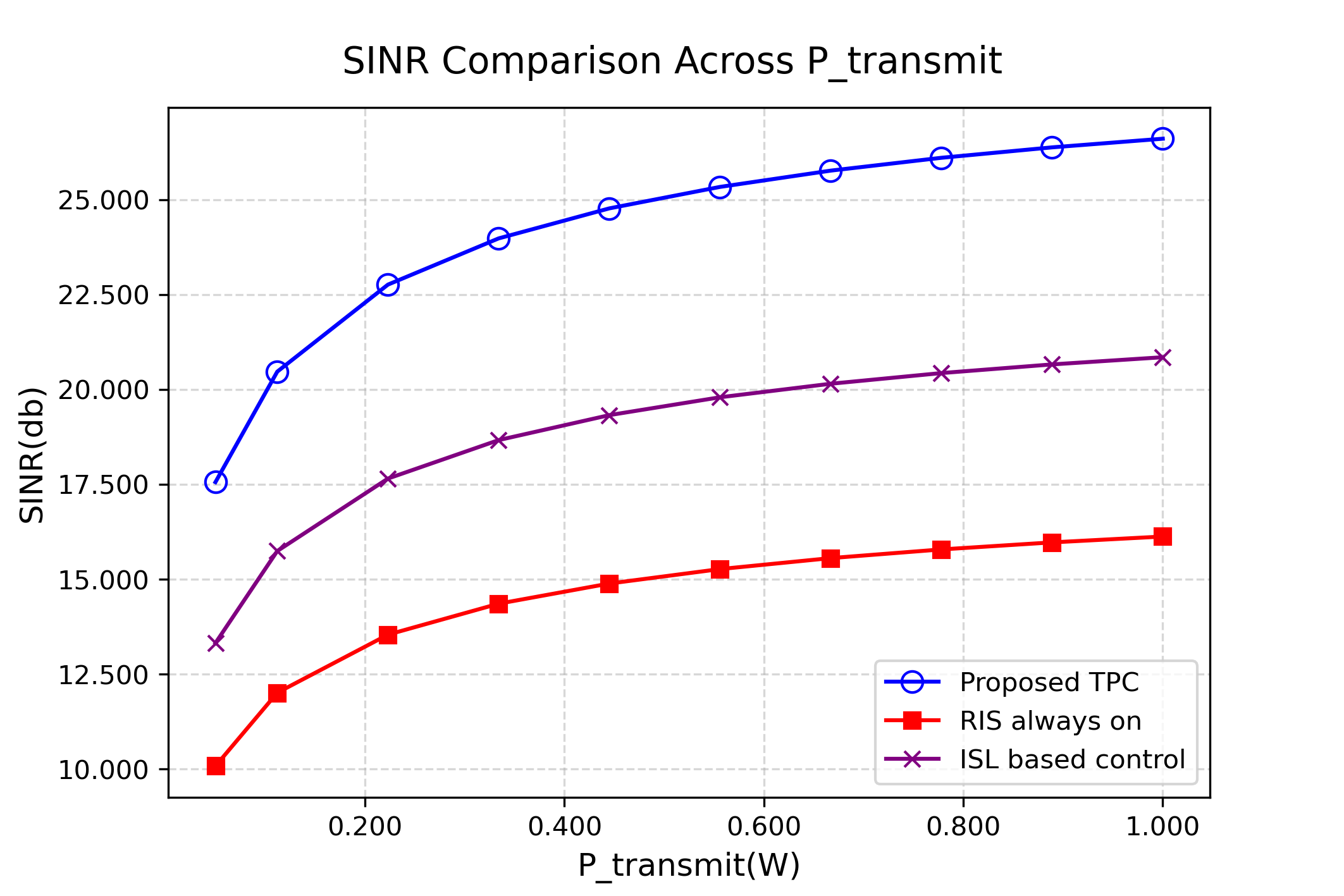}
    \caption{SINR comparison across transmission power}
    \label{fig:result1}
\end{figure}

\subsubsection{Simulation results}

\begin{figure}
    \centering
    \includegraphics[width=1.1\columnwidth]{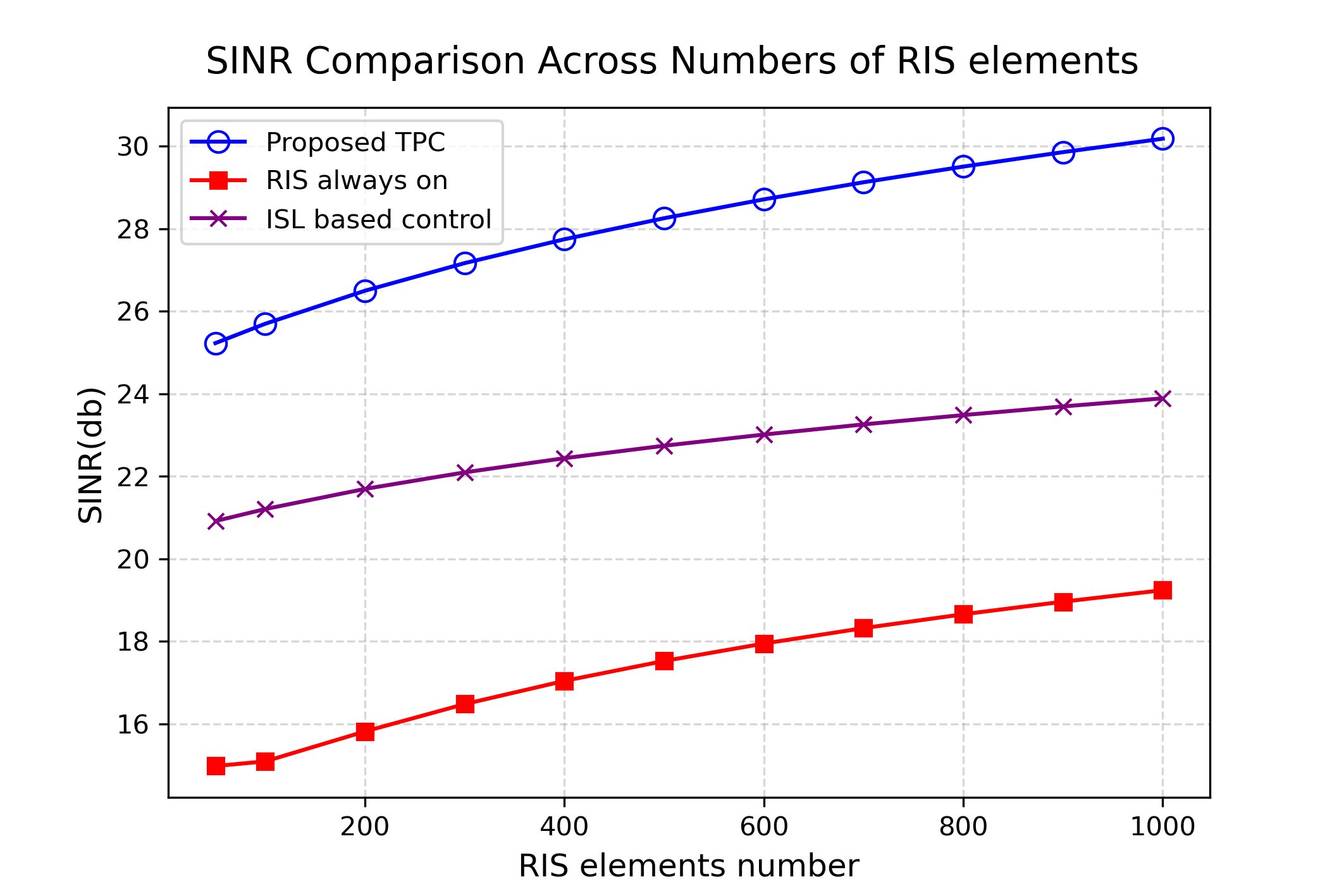}
    \caption{SINR comparison across numbers of RIS elements}
    \label{fig:result2}
\end{figure}

In this study, we compare our proposed method with two baseline approaches by varying the transmission power and configuring the RISs with different numbers of elements, aiming to demonstrate the advantages of our method clearly. The baseline approaches include the always-on RIS mode and the ISL-based control method \cite{tvt}.

For the proposed TPC method, we begin by identifying a known user trajectory from the dataset and apply a trajectory prediction model to forecast the user's trajectory over the next $50$ seconds. Similarly, the trajectories of $10$ interfering users can also be obtained. Then, we can calculate and compare the SINR of the user's transmitted signal, considering different transmission power levels and varying numbers of RIS elements. 



As shown in Fig.~\ref{fig:result1}, under different transmission powers, the SINR of all the three methods increases as expected. 
Notably, the proposed TPC method can predict the coordinates of users to determine the RIS ON-OFF states in advance, thereby mitigating interferences in near-real-time and thus the highest SINR can be achieved. In contrast, the ISL-based control scheme determines the current RIS ON-OFF state without predicting the future interference, so it cannot be adapted to the dynamic environment. Moreover, the method that `RIS always ON' shows the worst performance because it cannot sense the surrounding interference at all. Therefore, considering the interference-intensive case, the strength of the interference signal may exceed the desired signal due to the `blind-reflection' of RISs.

Similarly, as shown in Fig.~\ref{fig:result2}, under varying numbers of RIS elements, the SINR gradually increases with the number of elements. Meanwhile, we observe that as the number of elements increases, the interference amplification of RIS becomes more significant, resulting in severe spectrum pollution. The proposed TPC algorithm avoids unnecessary amplification of interference and achieves an optimal SINR.

\section{Conclusion }\label{s5}
In this paper, we present a trajectory prediction-based algorithm for mitigating interference in a dynamic uplink environment. Simulation experiments demonstrate that the proposed TPC scheme can significantly improve the achieved SINR by controlling the RIS ON-OFF state sequence in advance.  In future work, we will analyze how the accuracy of trajectory predictions impacts RIS control performance and explore effective strategies to address this challenge.

\bibliographystyle{IEEEtran}
\bibliography{IEEEabrv,reference}

\end{document}